\documentclass[12pt]{article}%
\usepackage{amsfonts}

\usepackage{bm}
\usepackage{cite}
\usepackage{mathrsfs}
\usepackage{amsmath}
\usepackage{graphicx}
\usepackage{verbatim}
\usepackage{color}
\usepackage{enumerate}
\usepackage{amsfonts}
\usepackage{amssymb}
\usepackage{authblk}
\usepackage{float}
\usepackage[hang,flushmargin]{footmisc}
\usepackage{lipsum}
\usepackage[font=footnotesize]{caption}
\usepackage{soul}
\usepackage{empheq}
\usepackage{tikz-cd}
\usepackage{makecell}
\usepackage{longtable}
\usepackage{framed}

\csname @addtoreset\endcsname{equation}{section}
\newcommand{\eal}[1]{\begin{equation} \begin{aligned} #1 \end{aligned}\end{equation}}

\usepackage[colorlinks=true,
            linkcolor=black,
            urlcolor=magenta,
            citecolor=blue]{hyperref}

 \newcommand{\badat}{\begin{alignedat}} 
 \newcommand{\eadat}{\end{alignedat}}
 
\def\bz{{\bar z}}
\def\Re{{\text{Re}}}

\definecolor{ForestGreen}{rgb}{0.13, 0.55, 0.13}
\definecolor{airforceblue}{rgb}{0.36, 0.54, 0.66}
\definecolor{orange}{rgb}{1.0, 0.5, 0.0}
\definecolor{amethyst}{rgb}{0.6, 0.4, 0.8}
\definecolor{awesome}{rgb}{1.0, 0.13, 0.32}
\definecolor{chromeyellow}{rgb}{1.0, 0.65, 0.0}

\begin{document}

\title{\vspace{-70pt} \Large{{\sc 
Celestial strings: field theory, conformally soft limits, and mapping the worldsheet onto the celestial sphere}
}\vspace{10pt}}
\author[a]{\normalsize{Lina Castiblanco}}
\author[b]{\normalsize{Gaston Giribet}}
\author[c]{\normalsize{Gabriel Marin}}
\author[d]{\normalsize{Francisco Rojas}}

\affil[a]{{\small\textit{School of Mathematics, Statistics and Physics, Newcastle University, Herschel Building, \protect\\ NE1 7RU Newcastle-upon-Tyne, U.K.}}}

\affil[b]{{\small\textit{Department of Physics, New York University. 726 Broadway, New York,
NY10003, USA.}}}

\affil[c]{{\small\textit{Departamento de F\'isica, FCFM, Universidad de Chile, Blanco Encalada 2008, Santiago, Chile.}}}

\affil[d]{{\small\textit{Facultad de Ingenier\'ia y Ciencias, Universidad Adolfo Ib\'a\~nez, Santiago, Chile.}}}

\date{}
\maketitle
\thispagestyle{empty}
\begin{abstract}
We compute the celestial correlators corresponding to tree-level 5-gluon amplitudes in the type I superstring theory. Since celestial correlation functions are obtained by integrating over the full range of energies, there is no obvious analog of the $\alpha' \to 0$ limit in this basis. This is manifestly shown by a factorization of the $\alpha '$ dependence in the celestial string amplitudes. Consequently, the question arises as to how the field theory limit is recovered from string theory in the celestial basis. This problem has been addressed in the literature for the case of 4-gluon amplitudes at tree level, where the forward scattering limit of the stringy factor was identified as a limit in which celestial Yang-Mills 4-point function is recovered. Here, we extend the analysis to the case with five gluons, for which the string moduli space allows for more types of limits, thus allowing to investigate this aspect in more detail. Based on celestial data only, we study the regime in which one arrives at the correct celestial field theory limit. We also study other properties of the celestial string amplitudes, namely, the conformally soft theorem, effective field theory expansion in the conformal basis, and a map that arises in the regime of high-energy/large-scaling dimension that connects the punctured string worldsheet to the insertion of primary operators in the celestial CFT for the massless $n$-point string amplitude. 
\end{abstract}

\newpage

\begin{small}
{\addtolength{\parskip}{-2pt}
 \tableofcontents}
\end{small}
\thispagestyle{empty}
\newpage

\section{Introduction}

The celestial holography program \cite{Strominger:2017zoo,deBoer:2003vf,Cheung:2016iub,Pasterski:2016qvg,Pasterski:2017kqt,Pasterski:2017ylz,Schreiber:2017jsr} is predominantly a bottom-up approach\footnote{However, see \cite{Costello:2022jpg, Costello:2022wso, Costello:2023hmi} for very interesting recent proposals for top-down constructions in the context of twisted holography.} to reformulate scattering amplitudes in asymptotically flat spacetimes in terms of correlators on a 2D CFT living on the Celestial Sphere, thus aiming to provide a holographic description of gravitational and gauge theories. The program gives a precise prescription to transform scattering amplitudes in asymptotically Minkowski space into correlation functions of a putative two-dimensional conformal field theory (CFT) on the celestial sphere, referred to as the celestial CFT (or CCFT for short). For massless particles, the precise relation between the observables of both theories is mediated by a Mellin transform. Concretely, the bulk scattering amplitudes of massless states in flat space can be expressed in the form of CCFT correlators by mapping the set of wave packets in the bulk into the Mellin transformed plane waves. This produces amplitudes of states that are eigenfunctions of the boost operators, with the eigenvalues translating into the conformal dimensions of dual primary operators. The action of the Lorentz group on constant-time sections of Minkowski null infinity represent M\"{o}bius transformations on the celestial sphere, and the operator product expansion (OPE) in the CCFT can be studied by analyzing the collinear limit of the amplitudes \cite{Pate:2019lpp}. A salient feature of the symmetry analysis in celestial holography is that the soft theorems of the bulk theory can be interpreted as Ward identities of the dual CFT corresponding to asymptotic symmetries that emerge near the conformal boundary of flat space \cite{Strominger:2013lka,Strominger:2013jfa}.

Despite many recent efforts to understand the nature of CCFTs from a large variety of points of view, there are still basic questions that remain open. For example, while there are strong reasons to think that CCFTs might be non-unitary, it is still unclear to which extent they resemble or differ from standard or non-unitary CFTs. For instance, there has been an ongoing discussion about whether the theory having a vanishing central charge. The current status of flat space holography certainly contrasts with the one of the AdS/CFT correspondence, where the pair of theories involved in the duality can be explicitly identified and the holographic dictionary accurately formulated. This is basically because in AdS/CFT one has access to a string theory formulation and, consequently, to a top-down derivation. One is thus immediately tempted to ask whether in the case of celestial holography a stringy realization is also possible. This is of course one important motivation to study celestial holography in the context of string theory. In relation to this point, it is natural to wonder whether there exists a celestial counterpart of the $1/N$ expansion of AdS/CFT. This question is related to the one alluded above regarding the fundamental structure of the CCFT since, for example, non-analytic dependencies in correlation functions, similar to those present in logarithmic CFTs \cite{Fiorucci:2023lpb,Agrawal:2023zea}, may appear in OPEs as an artifact of a perturbative truncation. All these are questions that, to be adequately addressed, would require a better knowledge of the theory.

Celestial holography was first studied in the context of string theory by Stieberger and Taylor in \cite{Stieberger:2018edy}, where tree-level scattering amplitudes in type I superstring theory and heterotic superstring were analyzed in the Mellin basis. Celestial tree-level amplitudes and OPEs were also studied in the context of string theories in \cite{Adamo:2019ipt, Casali:2020uvr, Adamo:2021zpw,Chang:2021wvv,Bu:2021avc, Guevara:2021tvr, Jiang:2021csc, Donnay:2023kvm}. In \cite{Stieberger:2018edy}, the authors studied the string scattering amplitudes in the celestial basis, focusing on the tree-level processes involving 4 gluons. One of the features observed when transforming the string amplitudes into the Mellin basis is that the dependence on $\alpha '$ factorizes out. {This factorization does not happen in other holographic proposals such as the Carrollian theory, cf. \cite{Stieberger:2024shv}.} In the case of celestial holography, the factorization of the dependence on $\alpha '$ turns out to be a direct consequence of the mix of energy regimes, and raises the question of how the field theory limit should be understood in this context. The authors of \cite{Stieberger:2018edy} made a proposal, showing that the string amplitudes do reproduce the correct limit at certain points of the moduli space: In contrast to what happens in string theory in momentum basis, where quantum field theory is recovered in the $\alpha'\to 0$ limit, in the celestial basis the field theory is recovered for the kinematic configurations that correspond to the forward scattering limit. This is arguably expected as in such a limit the processes turn out to be dominated by the exchanges of massless states. At the same time, this raises the question of what are exactly all the corners of the moduli space in which celestial string amplitudes reproduce the proper field theory limits. Motivated by this question, in \cite{Donnay:2023kvm} the authors continued the study of string celestial holography by exploring the structure of string amplitudes at 1-loop in the Mellin basis. Focusing on the case of 4-gluon processes, it was shown how the field theory limit of string theory does commute with the Mellin machinery that mixes the IR and UV regimes, which makes the analysis of the limit quite tractable. It was then observed that, as in the case of tree-level amplitudes, the 1-loop amplitudes, when translated into the celestial basis, also exhibit a simple dependence on $\alpha '$, given by a simple overall factor; consequently, the correct field theory limit involves a careful analysis of the different limits in the moduli space. Here, interested as we are in further exploring the moduli space of celestial string amplitudes, we will go back to the problem of tree-level processes and consider five gluon scattering amplitudes. 

The paper is organized as follows: In section 2, we review the celestial amplitudes in type I superstring theory. We discuss the 4- and 5-gluon scattering amplitudes both in momentum and in celestial basis. In section 3, we study the field theory limit. That is to say, we investigate the limits in which the celestial string amplitudes reduce to the YM analogs. In section 4, we discuss the conformally soft theorem for the case of the 5-gluons in the type I superstring. In Section 5 we investigate a map between the string worldsheet CFT and the celestial CFT that seems to emerge in the limit of high-energy/high-scaling-dimension for a general $n$-point string amplitude in momentum/conformal basis respectively. Section 6 contains a few remarks regarding the effective field theory expansion of string amplitudes in the conformal basis, and we finalize with closing words for future work.

\section{Celestial amplitudes for type I superstrings}

\subsection{Preliminary}

Let us discuss the 4- and 5-gluon celestial scattering amplitudes for type I superstring theory. The $n$-point celestial amplitude $\widetilde{\mathcal A}_n(\{\Delta_j,\vec{z}_j\})$ is obtained from the momentum space amplitude $\mathcal{A}_n(\{p_j\})$ by  evaluating the Mellin transform with respect to the external frequencies \cite{Pasterski:2016qvg,Pasterski:2017kqt,Pasterski:2017ylz,Schreiber:2017jsr}; namely
\eal{
\label{eq:Mellin}
\widetilde{\mathcal A}_{n}(\{\Delta_j,\vec{z}_j\}) &= \left(\prod_{j=1}^n \int_0^\infty \omega_j^{\Delta_j-1} d\omega_j\right) \delta^{(D)}\left(\sum_{j=1}^4 p_j^\mu\right)\mathcal{A}_n(\{p_j\}),
}
where $\Delta_{j} = d/2+i\lambda_{j}$ is the scaling dimension. The $D$-dimensional momentum $p^{\mu }_j$ of the $j^{\text{th}}$ external gluon ($j=1,2,...,n$; $\mu =0,1,...,D-1$) is parameterized in the celestial basis in terms of the corresponding positive frequency $\omega_j$ and the complex coordinates $\vec{z}_j$ of a point on the $d=D-2$ dimensional celestial sphere as follows 
\begin{equation}
\label{eq:celestial-p}
p_j^\mu (\omega_j,\vec{z}_j) = \eta_j \omega_j q^\mu(\vec{z}_j)\,,
\end{equation}
where $\eta_j=\pm 1$ represents the outgoing and incoming particles, respectively; and where 
\begin{equation}
q^\mu(\vec{z}_j) = (1+|\vec{z}_j|^2, 2\vec{z}_j,1-|\vec{z}_j|^2)\,,
\end{equation}
is a null vector pointing towards the celestial sphere. In $d=D-2=2$ dimensions, we use complex variables $\vec{z}_j=(z_j,\bar{z}_j)$.

In this work, we study the scattering of five gluons with momenta ${p_1,\dots,p_5}$, in the maximally helicity violating (MHV) configuration, where two gluons carry negative helicities, while the rest carry positive helicities. For the MHV configuration, the tree-level scattering amplitude for $n=4$ and $n=5$ gluons, in the type I superstring ($\mathcal{A}_{I}^{\text{tree}}$), can be written as the product of the color-ordered YM amplitude ($\mathcal{A}_{\text{YM}}^{\text{tree}}$) and the type I string form factor (${F}_{I}^{\text{tree}}$) \cite{Stieberger:2006bh}; that is,
\begin{equation}
    \mathcal{A}_{I}^{\text{tree}} (-,-,+,\dots,+)\,=\, \mathcal{A}_{\text{YM}}^{\rm tree} (-,-,+,\dots,+) \, {F}_{I}^{\text{tree}}(\{s_{ij}\})\,.
\end{equation}
This factorization has been shown to hold for the cases $n=4$ and $n=5$ \cite{Stieberger:2006bh}, and it is believed to be a general feature for higher points (see \emph{e.g.}, \cite{Mafra:2011nv}). The YM amplitude contains all the helicity information of the colliding gluons, while the form factor ${F}_{I}^{\text{tree}}$ encodes the stringy structure in terms of the Mandelstam invariants only, defined as\footnote{In this article we use the mostly-plus metric $\eta_{\mu\nu}=diag(-,+,\dots,+)$.}
\begin{equation}
    s_{ij}  = -\left(p_i+p_{j}\right)^2 =  \eta_i\eta_{j}\omega_{i}\omega_{j}z_{ij}\bar{z}_{ij}\,,
    \label{eq:k-invariants}
\end{equation}
with $z_{ij}\equiv z_i - z_j$. The  color-ordered YM amplitude is given by \cite{Parke:1986gb,Berends:1987me}
\begin{equation}
    \mathcal{A}_{\text{YM}}^{\rm tree} (-,-,+,\dots,+) = ig^{n-2}\,\text{Tr}(T^1\cdots T^n)\,\dfrac{\langle 12\rangle^{4}}{\langle 12\rangle \langle 23\rangle \cdots \langle n1\rangle}\,,
    \label{Aym}
\end{equation}
where the $T^a$ are the algebra generators in the fundamental representation of the gauge group; $g$ is the YM coupling constant. Also, using the parametrization \eqref{eq:celestial-p} for the momenta, the spinor products become
\begin{equation}
    \langle ij \rangle = \sqrt{\omega_{i}\omega_{j}}z_{ij}\bar{z}_{ij}\,, \qquad [ ij ] = - \sqrt{\omega_{i}\omega_{j}}z_{ij}\bar{z}_{ij}\,.
\end{equation}

\subsection{4-gluon superstring amplitude}
Let us start by reviewing the 4-gluon scattering amplitude in the type-I theory in both, momentum basis \cite{Green:1982sw, Stieberger:2006bh}, and celestial basis \cite{Stieberger:2018edy, Donnay:2023kvm}.

\subsubsection{4-gluon string amplitude in momentum basis}

At tree-level, the MHV string amplitude with four external gluons reads
\eal{
\label{4gluon-string}
\mathcal{A}_{I}^{\text{tree}} (-,-,+,+)= \mathcal{A}_{\text{YM}}^{\rm tree} (-,-,+,+) F_{I}^{\text{tree}}(s,t)\,,
}
where the color-ordered YM 4-gluon amplitude is given by
\eal{
\label{eq:4MHV}
\mathcal{A}_{\rm YM}^{\rm tree} (-,-,+,+) = \frac{\omega_1 \omega_2}{\omega_3 \omega_4} \frac{z_{12}^3}{z_{23} z_{34} z_{41}} = r \frac{z_{12}\bar z_{34}}{\bar z_{12} z_{34}} \,,
}
where $r$ is the conformally invariant cross ratio 
\eal{
\label{r-angle}
r= \frac{z_{12}z_{34}}{z_{23}z_{41}}\,.
}
In the last equality of \eqref{eq:4MHV} we have used the total momentum conservation, which cancels out all the energy factors $\omega_i$ from it.

Our convention for the Mandelstam variables will be 
\begin{align}
s & =-(p_1+p_2)^2\,,\\
t & = -(p_2+p_3)^2\,, \\
u & =-(p_1+p_3)^2\,.    
\end{align}
The invariant cross-ratio and Maldelstam variables are related to the scattering angle $\theta$ in the centre-of-mass frame through 
\eal{
\label{eq:cross-ratio}
r =-\frac{s}{t} = \csc^2\left(\frac{\theta}{2}\right).
}
For 4-gluons, the tree-level string form factor is written in terms of the Euler beta function $B(x,y)$, namely
\eal{
\label{eq:Ftree}
F_I^{\rm tree}(s,t) = -\alpha' s  B(-\alpha' s,1+\alpha' t)\,.
}

The scattering angle $\theta$ is crucial in the discussion of the field theory limit in the celestial basis \cite{Stieberger:2018onx}. We review the amplitudes in the celestial basis in the next subsection.

\subsubsection{4-gluon string amplitude in celestial basis}

Here we review \cite{Stieberger:2018edy}. The celestial amplitude is obtained by Mellin transforming  \eqref{4gluon-string}; namely, 
\eal{
\tilde{\mathcal{A}}_I^{\rm tree}(\lambda_i,z_i,\bar z_i) &= \left( \prod_{i=1}^4 \int_0^\infty d\omega_i \,  \omega_i^{i\lambda_i}\right)  \delta^{4}(\omega_1 q_1 + \omega_2 q_2 - \omega_3 q_3 -\omega_4 q_4) \,  r \frac{z_{12}\bar z_{34}}{\bar z_{12} z_{34}}\, F_I^{tree}(s,u).
}
Expressing the total momentum conserving $\delta $-function  as
\eal{
\label{eq:4-delta}
&\delta^{(4)}(\omega_1 q_1 + \omega_2 q_2 - \omega_3 q_3 -\omega_4 q_4) = \frac{4}{\omega_4|z_{14}|^2 |z_{23}|^2} \delta(r-\bar r)\\
& \hspace{100pt} \times \delta\left(\omega_1-\frac{z_{24}\bar{z}_{34}}{z_{12}\bar{z}_{13}}\omega_4\right)  \delta\left(\omega_2-\frac{z_{14}\bar{z}_{34}}{z_{12}\bar{z}_{32}}\omega_4\right)  \delta\left(\omega_3-\frac{z_{42}\bar{z}_{14}}{z_{23}\bar{z}_{13}}\omega_4\right)\,,
}
it is possible to eliminate (localize) three integrals, obtaining  
\begin{multline}
    \tilde{\mathcal{A}}_I^{\rm tree}(\lambda_i,z_i,\bar z_i) =  \frac{4r^3}{\bar z_{12}^2 z_{34}^2} \delta(r-\bar r) \left(\frac{\bar z_{34}}{z_{12}}\right)^{i \lambda_1} \left(\frac{z_{34}}{\bar z_{12}}\right)^{i \lambda_2}  \left(\frac{z_{24}}{\bar z_{13}}\right)^{i (\lambda_1+\lambda_3)} \left(\frac{\bar z_{14}}{z_{23}}\right)^{i (\lambda_2+\lambda_3)} 
    \\
     \times \Theta(r-1)  \int_0^\infty \omega_4^{2i\beta-1} F_I^{\rm tree }(s,u)\,d\omega_4\,, 
\end{multline}
where the step function $\Theta(r-1)$ simply enforces the condition for physical scattering $r>1$, and $\beta \equiv -\frac{i}{2}\sum_{i=1}^4 \lambda_i$. 
One can further simplify the last equation by using the new integration variable $w\equiv s/r$ such that 
\eal{
\omega_4^2 = \frac{r|z_{13}|^2 w}{(r-1)|z_{14}|^2|z_{34}|^2}\,,
}
and the tree-level open string amplitude in the celestial basis becomes
\eal{
\label{eq:tree-cstring}
\tilde{\mathcal{A}}_I^{\rm tree}(\lambda_i,z_i,\bar z_i) &= 4 \delta(r-\bar r) \left(\frac{\bar z_{34}}{z_{12}}\right)^{i \lambda_1} \left(\frac{z_{34}}{\bar z_{12}}\right)^{i \lambda_2}  \left(\frac{z_{24}}{\bar z_{13}}\right)^{i (\lambda_1+\lambda_3)} \left(\frac{\bar z_{14}}{z_{23}}\right)^{i (\lambda_2+\lambda_3)} \\
&\times \Theta(r-1) \frac{r^3}{\bar z_{12}^2 z_{34}^2} \left(\frac{r|z_{13}|^2}{(r-1)|z_{14}|^2|z_{34}|^2}\right)^{-\beta} I(r,\beta)\,,
}
where
\eal{
\label{Iintegral}
I(r,\beta) = \frac{1}{2} \int_0^\infty w^{-\beta-1}F_I^{\rm tree}(rw,-w) \, dw\,.
}
This can be rewritten as the 4-point CFT correlator
\eal{
\label{eq:4pt-correlator-tree}
\tilde{\mathcal{A}}_I^{\rm tree}(\lambda_i,z_i,\bar z_i) &= f(r,\bar r) \prod_{i<j}^4 z_{ji}^{\frac{h}{3}-h_i-h_j} \bar z_{ji}^{\frac{\bar h}{3}-\bar h_i-\bar h_j}\,, 
}
with $h=\sum_{i=1}^4 h_i $, $\bar h=\sum_{i=1}^4 \bar h_i$, together with the weights 
\eal{\label{eq:weights}
h_1 &= \frac{i}{2}\lambda_1\,,\quad \quad &  \bar  h_1 =1+\frac{i}{2}\lambda_1\,, \\
h_2 &= \frac{i}{2}\lambda_2\,, \quad \quad & \bar h_2 =1+\frac{i}{2}\lambda_2\,, \\
h_3 &= 1+\frac{i}{2}\lambda_3\,, \quad &  \bar h_3=\frac{i}{2}\lambda_3\,, \\
h_4 &= 1+\frac{i}{2}\lambda_4\,, \quad &  \bar h_4 =\frac{i}{2}\lambda_4\,.\\
}
and with $f(r,\bar r)$ being a function that only depends on the conformally invariant cross-ratios $(r,\bar r)$. Explicitly,
\eal{
\label{eq:f}
f(r,\bar r) = 4(\alpha')^\beta \delta(r-\bar r) \Theta(r-1) r^{\frac{5-\beta}{3}}(r-1)^{\frac{2-\beta}{3}}I(r,\beta)\,.
}
The careful analysis of this function is required to explore the limits in which the string theory amplitude of 4 gluons in the celestial basis yields the YM analog. As said, this has been explored in detail for 4 gluons in \cite{Stieberger:2018edy} and\cite{Donnay:2023kvm}. Here, we are going to explore the case of five gluons whose moduli space exhibits more structure and, therefore, more limits are possible.

\subsection{5-gluon superstring amplitude}

We now consider the 5-gluon string amplitude in both bases. Since the number of independent Mandelstam invariants for an $n$-point amplitude is \mbox{$3n\!-\!10$}, for the $n=5$ case it suffices to define five invariants which, according to \eqref{eq:k-invariants}, can be conveniently chosen to be $s_{i,i+1}\equiv s_i$, with the identification $i+n \sim i$. 

We will start reviewing the known literature of this amplitude in momentum basis.

\subsubsection{The 5-gluon string amplitude in momentum basis}
The 5-gluon MHV scattering amplitude in type I superstring theory is given by \cite{Stieberger:2006bh,Stieberger:2006te}
\begin{equation}
	\label{5gluon-string}
	\mathcal{A}_{I}^{\rm tree} (-,-,+,+,+)= \mathcal{A}_{\rm YM}^{\rm tree} (-,-,+,+,+) \left[V^{(5)}(s_i)-2iP^{(5)}(s_i)\epsilon(1,2,3,4)\right],
\end{equation}
with
\begin{equation}
    P^{(5)}(s_i) = \int^{1}_{0}\,dx\int^{1}_{0}\,dy\, x^{\alpha's_2}y^{\alpha's_5}(1-x)^{\alpha's_3}(1-y)^{\alpha's_4}(1-xy)^{\alpha'(s_1-s_3-s_4)-1}\,,
    \label{eq:p5}
\end{equation}
\begin{multline}
      V^{(5)}(s_i) = \alpha'^2s_2s_5\int^{1}_{0}\,dx\int^{1}_{0}\,dy\, x^{\alpha's_2-1}y^{\alpha's_5-1}(1-x)^{\alpha's_3}(1-y)^{\alpha's_4}(1-xy)^{\alpha'(s_1-s_3-s_4)} \\+ \tfrac{1}{2}\alpha'^2\left(s_2s_3+s_4s_5-s_1s_2-s_3s_4-s_1s_5\right)P^{(5)}(s_i)\,,
      \label{eq:v5}
\end{multline}
and 
\begin{equation}
\epsilon(1,2,3,4) = \alpha'^2\epsilon_{\alpha\beta\mu\nu}p_1^{\alpha}p_2^{\beta}p_3^{\mu}p_4^{\nu} = \alpha'^{2}\prod_{j=1}^{4} \eta_{j}\omega_{j}\epsilon_{\alpha\beta\mu\nu}q_{1}^{\alpha}q_{2}^{\beta}q_{3}^{\mu}q_{4}^{\nu} \,.
\end{equation}
The color-ordered YM 5-gluon amplitude can be expressed as 
\begin{equation}
    \mathcal{A}_{\rm YM}^{\rm tree}({-,-,+,+,+}) = \frac{i}{4}\frac{(1-r_4)(1-\bar{r}_4)}{r_4-\bar{r}_4} \frac{f_{15}f_{25}}{\omega_5f_{35}f_{45}} \frac{z_{12}^3}{z_{23}z_{34}z_{45}z_{51}|z_{14}|^2|z_{23}|^2}\prod_{i=1}^4\delta(\omega_i-\omega^*_i)\,.
    \label{5-point-aym}
\end{equation}
In the last line, we used the $5$-point momentum $\delta$-function given by \cite{Fan:2019emx}
\begin{equation}
\label{delta-func}
    \delta^{(4)}\left( \sum_{i=1}^{5}\eta_{i}\omega_{i}q_{i} \right) = \frac{i}{4}\frac{(1-r_4)(1-\bar{r}_4)}{r_4-\bar{r}_4}\frac{1}{|z_{14}|^2|z_{23}|^2} \prod_{i=1}^4\delta(\omega_i-\omega^*_i)\,,
\end{equation}
where $\omega^*_i=f_{i5}\,\omega_5$, and $f_{i5}$ are functions of the celestial coordinates $(z_i , \bar{z}_i )$ defined as follows
\begin{align}
    f_{15} &= \eta_1\eta_5r_4\left|\frac{z_{24}}{z_{12}}\right|^2\frac{(1-r_4)(1-\bar{r}_4)}{r_4-\bar{r}_4}\frac{r_5-\bar{r}_5}{(1-r_5)(1-\bar{r}_5)}\left|\frac{z_{15}}{z_{14}}\right|^2 - \eta_1\eta_5 r_5\left|\frac{z_{25}}{z_{12}}\right|^2\,, \\
    f_{25} &= -\frac{\eta_1\eta_5}{\eta_1\eta_2}\frac{1-r_4}{r_4} \left|\frac{z_{34}}{z_{23}}\right|^2\frac{(1-r_4)(1-\bar{r}_4)}{r_4-\bar{r}_4}\frac{r_5-\bar{r}_5}{(1-r_5)(1-\bar{r}_5)}\left|\frac{z_{15}}{z_{14}}\right|^2 +\frac{\eta_1\eta_5}{\eta_1\eta_2} \frac{1-r_5}{r_5}\left|\frac{z_{35}}{z_{23}}\right|^2\,,\\
    f_{35} &=\frac{\eta_1\eta_5}{\eta_1\eta_3}(1-r_4) \left|\frac{z_{24}}{z_{23}}\right|^2\frac{(1-r_4)(1-\bar{r}_4)}{r_4-\bar{r}_4}\frac{r_5-\bar{r}_5}{(1-r_5)(1-\bar{r}_5)}\left|\frac{z_{15}}{z_{14}}\right|^2-\frac{\eta_1\eta_5}{\eta_1\eta_3}(1-r_5)\left|\frac{z_{25}}{z_{23}}\right|^2\,,\\
    f_{45} &=-\frac{\eta_1\eta_5}{\eta_1\eta_4}\frac{(1-r_4)(1-\bar{r}_4)}{r_4-\bar{r}_4}\frac{r_5-\bar{r}_5}{(1-r_5)(1-\bar{r}_5)}\left|\frac{z_{15}}{z_{14}}\right|^2\,,
\end{align}
and the two cross-ratios suitable for five particles,
\begin{equation}
    r_4 = \frac{z_{12}z_{34}}{z_{13}z_{24}}\,,\quad r_5 = \frac{z_{12}z_{35}}{z_{13}z_{25}}\,.
 \end{equation}
Therefore, the 5-gluon superstring amplitude in momentum basis reads
\begin{equation}
    \mathcal{A}_{I}^{tree} (-,-,+,+,+)= \frac{i}{4}\frac{(1-r_4)(1-\bar{r}_4)}{r_4-\bar{r}_4} \frac{f_{15}f_{25}}{f_{35}f_{45}} \frac{z_{12}^3}{z_{23}z_{34}z_{45}z_{51}|z_{14}|^2|z_{23}|^2}\omega_5^{-1} \nonumber 
    \end{equation}
    \begin{equation}
 \ \ \ \ \ \ \ \ \ \    \times \left[V^{(5)}-2iP^{(5)}\epsilon(1,2,3,4)\right](\omega_{i},z_{i})\,.    \label{5gluon-string-mb}  
\end{equation}
In the next section, we will derive the celestial counterpart of this expression.

\subsubsection{The 5-gluon string amplitude in celestial basis}

To translate the 5-gluon superstring amplitude from the momentum basis to the celestial one, we perform the Mellin transformation of equation \eqref{5gluon-string-mb} as in \eqref{eq:Mellin}; namely
\begin{equation}
\tilde{\mathcal{A}}_I^{\rm tree}(\lambda_i,z_i,\bar z_i) =  \left(\prod_{j=1}^5 \int_0^\infty \omega_j^{i\lambda_j} d\omega_j\right) \delta^{(4)}\left(\sum_{i=1}^{5}\eta_{i}\omega_{i}q_{i}\right)\mathcal{A}_I^{\rm tree}(\{\omega_j, \vec{z}_j\})\,.
\end{equation}
Thus, by employing the $\delta$-function we can eliminate all integrals except the one on $\omega_{5}$. This results in the following expression
\begin{multline}
\label{eq:omega5int}
    \tilde{\mathcal{A}}_I^{\rm tree}(\lambda_i,z_i,\bar z_i) = \frac{i}{4} \frac{(1-r_4)(1-\bar{r}_4)}{r_4-\bar{r}_4}\frac{z_{12}^3}{|z_{14}|^2|z_{23}|^2z_{23}z_{34}z_{45}z_{51}} \\ \times f_{15}^{i\lambda_1+1}f_{25}^{i\lambda_2+1}f_{35}^{i\lambda_3-1}f_{45}^{i\lambda_4-1}
    \int^{\infty}_{0}\,d\omega_{5}\, \omega_5 ^{-2\beta-1} \left[V^{(5)}(s_i)-2iP^{(5)}(s_i)\epsilon(1,2,3,4)\right] \,,
\end{multline}
where now $\beta = -\frac{i}{2}\sum_{i=1}^{5}\lambda_i$. In order to perform the integral above, we need to find out how $\omega_{5}$ depends on the 5 kinematic invariants. This can be determined by using (\ref{delta-func}). This yields
\begin{gather} 
    s_1  = \eta_{1} \eta_{2} f_{15}f_{25}|z_{12}|^2 \omega_5^2\,, \qquad 
    s_2  =  \eta_{2} \eta_{3}f_{25}f_{35}|z_{23}|^2\omega_5^2\,, \qquad 
    s_3  =  \eta_{3} \eta_{4} f_{35}f_{45}|z_{34}|^2 \omega_5^2\,, \nonumber \\
    s_4  =  \eta_{4} \eta_{5}f_{45}|z_{45}|^2\omega_5^2\,, \qquad
    s_5  =  \eta_{1}\eta_{5} f_{15}|z_{51}|^2 \omega_5^2\,.
\end{gather}
Similarly, for the pseudoscalar we have
\begin{equation}
    \epsilon(1,2,3,4)  = \alpha'^2 \omega_5^4 \eta_{1}\eta_{2}\eta_{3}\eta_{4}f_{15}f_{25}f_{35}f_{45}\epsilon_{\alpha\beta\mu\nu}q_1^{\alpha}q_2^{\beta}q_3^{\mu}q_4^{\nu}  \equiv \alpha'^2 \omega_5^4 c_3 \,. \label{c3}
\end{equation}
At this stage, it is convenient to write $P^{(5)}(s_{i})$ and $V^{(5)}(s_{i})$ as follows
\begin{align}
    P^{(5)}(s_i) & = \int^{1}_{0}\,dx\int^{1}_{0}\, dy\, (1-xy)^{-1} e^{-\alpha' A(z_i,\bar{z}_i) \omega_5^2}\,,\\
    V^{(5)}(s_i) & = \alpha'^2 \left[c_1\int^{1}_{0}\,dx\int^{1}_{0}\, dy\, (xy)^{-1} e^{-\alpha' A(z_i,\bar{z}_i) \omega_5^2} + c_2 P^{(5)}(s_i)\right]\omega_5^4\,,
\end{align}
where
\begin{multline}
    A(z_i,\bar{z}_i) = -\big[ \eta_{2}\eta_{3}f_{25}f_{35}|z_{23}|^2 \log x +\eta_{5}\eta_{1}f_{15}|z_{51}|^2 \log y + \eta_{3}\eta_{4}f_{35}f_{45}|z_{34}|^2 \log(1-x) \\ + \eta_{4}\eta_{5}f_{45}|z_{45}|^2 \log (1-y) 
    + (\eta_{1}\eta_{2}f_{15}f_{25}|z_{12}|^2-\eta_{3}\eta_{4}f_{35}f_{45}|z_{34}|^2-\eta_{4}\eta_{5}f_{45}|z_{45}|^2) \log(1-xy)\big]\,,
    \label{mellin-string-A}
\end{multline}
and where the coefficients $c_{1}$ and $c_{2}$, which depend on $(z_i, \bar{z}_i)$, are
\begin{align} 
    c_1 & = \eta_{1}\eta_{2}\eta_{3}\eta_{5}f_{25}f_{35}f_{15}|z_{23}|^2|z_{51}|^2 \label{c1}\,,\\ 
    c_2 &= \tfrac{1}{2}\left(\eta_{2}\eta_{4}f_{25}f_{35}^2f_{45}|z_{23}|^2 |z_{34}|^2 + \eta_{1}\eta_{4}f_{45}f_{15}|z_{45}|^2 |z_{51}|^2 \right. \nonumber\\
    &\left. \qquad - \eta_{1}\eta_{3}f_{15}f_{25}^2f_{35}|z_{12}|^2 |z_{23}|^2 - \eta_{3}\eta_{5}f_{35}f_{45}^2|z_{34}|^2 |z_{45}|^2-\eta_{2}\eta_{5}f_{15}^2f_{25}|z_{12}|^2 |z_{51}|^2\right)\,. \label{c2}
\end{align}
By using the previous definitions, the $\omega_5$ integral in (\ref{eq:omega5int}) is shown to yield
\begin{multline}
\label{eq:w5-integral}
    \int^{\infty}_{0}\, \omega_5 ^{-2\beta-1} \left[V^{(5)}(s_i)-2iP^{(5)}(s_i)\epsilon(1,2,3,4)\right]d\omega_{5} \\ = \tfrac{1}{2}\alpha'^{\beta}\Gamma(2-\beta)\int^{1}_0\, dx \int^{1}_0\, dy \left[c_1(xy)^{-1}+(c_2-2ic_3)(1-xy)^{-1}\right]A^{\beta-2} \,,
\end{multline}
with $\Re(\beta)<2$ and $\Re(A)>0$. Therefore, the tree-level 5-gluon scattering amplitude in type I superstring theory written in celestial basis takes the following form 
\begin{multline}
\label{eq:5pointstring-st}
    \tilde{\mathcal{A}}_I^{\rm tree}(\lambda_i,z_i,\bar z_i) = i\frac{\alpha'^{\beta}}{8}\Gamma(2-\beta)\frac{(1-r_4)(1-\bar{r}_4)}{r_4-\bar{r}_4}\frac{z_{12}^3}{|z_{14}|^2|z_{23}|^2z_{23}z_{34}z_{45}z_{51}} \\ \times f_{15}^{i\lambda_1+1}f_{25}^{i\lambda_2+1}f_{35}^{i\lambda_3-1}f_{45}^{i\lambda_4-1}
    \int^{1}_0\, dx \int^{1}_0\, dy \left[c_1(xy)^{-1}+(c_2-2ic_3)(1-xy)^{-1}\right]A^{\beta-2} \,,
\end{multline}
with $c_1, c_2, c_3$ and $A$ given above, in equations \eqref{c1}, \eqref{c2}, \eqref{c3} and \eqref{mellin-string-A}, respectively. Note that, although we needed the condition $\Re(\beta)<2$ in order for the energy integral \eqref{eq:w5-integral} to converge, the resulting gamma function $\Gamma(2-\beta)$ defines the analytic continuation to the entire complex-$\beta$ plane.

One remarkable property of the expression (\ref{eq:5pointstring-st}) is that its dependence of $\alpha'$ factorizes out. In the calculation, this occurs when performing the Mellin transformation, \emph{i.e.}, it is a consequence of the integration over all the energies when going to the celestial basis. The same factorization phenomenon has been observed in \cite{Stieberger:2018edy} and \cite{Donnay:2023kvm} for the 4-gluon string amplitude, both at three-level and at one-loop. It is worth emphasizing that this is a peculiar feature of celestial amplitudes and does not happen, for example, in the Carrollian amplitudes \cite{Stieberger:2024shv}. Here, we observe that the same factorization takes place in the celestial 5-gluon string scattering at tree level, and, on dimensional grounds, it is expected to occur in all string amplitudes for any multiplicity and loop level. 

In the next section, we will study other properties of formula (\ref{eq:5pointstring-st}). More precisely, we will explore the limits in which this reproduces the field theory result in the celestial basis.

\section{The celestial field theory limit}

The observation that the $\alpha' \to 0$ limit of string amplitudes reproduces field theory ones, goes back to the early days of the dual resonance models. By taking the $\alpha'\to 0$ limit of the original Veneziano amplitude \cite{Veneziano:1968yb}, Scherk noticed that holding both, the mass of the tachyon and the ratio $g_s/\sqrt{\alpha'}\equiv \lambda$ fixed, while taking the limit $\alpha ' \to 0$, yielded the tree-level amplitude of massive scalars in the $\lambda \varphi^3$ theory \cite{Scherk:1971xy}. Later on, by including SU($N$) Chan-Paton degrees of freedom attached at open string end-points, Neveu and Scherk showed that this string model was able to reproduce the tree-level gluon amplitude in Yang-Mills theory \cite{Neveu:1971mu}. Superstring amplitudes are certainly not the exception to this; moreover, since the lowest energy states of the superstring spectrum are massless states, their low energy limit is not plagued by the usual ambiguities that the tachyon poses in the bosonic string theory.

As pointed out at the end of the previous section, the $\alpha'$ dependence in celestial string amplitudes is a simple overall factor. Then, this begs the question of how to recover the field theory limit from the string amplitude in the conformal basis, that is, 
\begin{figure}[h!]
\begin{center}
    \begin{tikzcd}[column sep=2cm, row sep=2cm]
    \makecell{\text{String}\\ \text{Amplitude}} \arrow[r, "\text{Mellin}"] \arrow[d, "\rotatebox{90}{$\alpha'\rightarrow 0$}"']
    & \makecell{\text{Celestial String}\\ \text{Amplitude}} \arrow[d, "\text{?}"] \\
    \makecell{\text{Field Theory}\\ \text{Amplitude}} \arrow[r, "\text{Mellin}"]
    & \makecell{\text{Celestial Field Theory}\\ \text{Amplitude.}}
    \end{tikzcd}
\end{center}
\end{figure}
\vspace{-0.6cm}

In this section we will first review how to do this for the 4-gluon case, and then we develop its extension to 5-gluons which is endowed with a richer kinematic structure.

\subsection{4-gluon limit}

In \cite{Stieberger:2018edy} Stieberger and Taylor have shown that the tree-level celestial 4-gluon field theory amplitude is recovered in the $r \to \infty$ (forward scattering) limit of the celestial string amplitude \eqref{eq:tree-cstring}\footnote{Using crossing symmetry, one can see that taking $r\to 0$ also recovers the correct field theory limit \cite{Donnay:2023kvm}.}. To see this, we first expand $I(r,\beta)$ in powers of $1/r$. Then, it is convenient to use the integral representation of the Euler beta function, that is
\eal{
\label{eq:4-point-integral}
F_I^{\rm tree}(rw,-w) = -rw \int_0^1 x^{-rw-1}(1-x)^w \, dx \,,
}
where, in order to facilitate the comparison with \cite{Stieberger:2018edy}, we have rescaled $w\to  w/\alpha'$. Inserting this into \eqref{Iintegral}, we have
\eal{
I(r,\beta) &= -\frac{r}{2}\int_0^\infty dw w^{-\beta} \int_0^1 dx\, x^{-rw-1}(1-x)^w   \\
&=-\frac{r}{2}\int_0^1 \frac{dx}{x}\,\int_0^\infty dw\, w^{-\beta} e^{-w\left[r\log x-\log(1-x)\right]}\,.
}
Defining $y\equiv w\left[r\log x-\log(1-x)\right]$, we have
\eal{
\label{eq:Itree}
I(r,\beta) &=-\frac{r}{2}\Gamma(1-\beta)\int_0^1 \frac{dx}{x}\,\left[r\log x-\log(1-x)\right]^{\beta-1}\,.
}
So far, this result is exact. If we now expand in powers of $1/r$, we obtain
\eal{
I(r,\beta) &= -\frac{1}{2}r^{\beta}\Gamma(1-\beta)\int_0^1 \frac{dx}{x}\,(\log x)^{\beta-1} +\mathcal{O}(r^{\beta-1})\,.
}
Making the change of variables $t\equiv \log x$, we end up with
\eal{
I(r,\beta) &=-\frac{1}{2}r^{\beta}\Gamma(1-\beta)\int_{-\infty}^0 dt \,t^{\beta-1} +\mathcal{O}(r^{\beta-1})\,.
}
Changing variables again, now defining $\text{z}\equiv \log(- t)$, we have
\eal{
I(r,\beta) &= \frac{1}{2}r^{\beta}\Gamma(1-\beta)\int_{-\infty}^{\infty} d\text{z} \,e^{\beta \text{z}}+\mathcal{O}(r^{\beta-1})\,,
}
and thus, 
\eal{
I(r,\beta) &= 2\pi \delta(\lambda_1+\lambda_2+\lambda_3+\lambda_4)+\mathcal{O}(r^{\beta-1})\,.
}
Therefore, the $r \to \infty$ limit of the celestial correlator corresponding to the tree-level 4-gluon string amplitude \eqref{eq:tree-cstring} reduces to
\eal{
\lim_{r \to \infty}\tilde{\mathcal{A}}_I^{\rm tree}(\lambda_i,z_i,\bar z_i) &=  8\pi \delta(r-\bar r) \left(\frac{\bar z_{34}}{z_{12}}\right)^{i \lambda_1} \left(\frac{z_{34}}{\bar z_{12}}\right)^{i \lambda_2}  \left(\frac{z_{24}}{\bar z_{13}}\right)^{i (\lambda_1+\lambda_3)} \left(\frac{\bar z_{14}}{z_{23}}\right)^{i (\lambda_2+\lambda_3)} \\
&\times \theta(r-1) \frac{r^3}{\bar z_{12}^2 z_{34}^2}   \delta(\lambda_1+\lambda_2+\lambda_3+\lambda_4)\,,
}
which is exactly the tree-level 4-gluon celestial amplitude in YM theory. In the case of 4 gluons this is understood as the limit in which the amplitude is dominated by the exchange of massless states. 

It is important to stress the fact that the forward limit ($r\to \infty$) above is only taken on the $I(r,\beta)$ function, \emph{i.e.}, on the stringy factor of the full celestial amplitude in \eqref{eq:f}, while the Yang-Mills 4-point CFT correlator in \eqref{eq:4pt-correlator-tree} remains intact. This prescription allows us to arrive at the correct field theory limit for an arbitrary kinematic configuration, not only in the forward scattering one. 

Let us now move to investigate what is the analog in the case of amplitudes that involve more gluons.

\subsection{5-gluon limit}

We aim at exploring further the limits in which the celestial string scattering amplitudes reproduce the YM theory results. More concretely, our intention is to extend the analysis reviewed above to the case $n=5$. This will provide us with a richer limiting structure in the celestial parameters ($z_i,\bar z_i$) to play with and, consequently, it will enable us to have a better picture of what are the specific corners in the space of kinematic variables where the field theory result is recovered. 

\subsubsection{Single Regge limit}

The complexity of the Regge limits increases for amplitudes where $n>4$ as, in such case, there are at least two limits that is worth distinguishing: the single-Regge limit and the multi-Regge limit, the former being of special interest as it is the direct generalization of the limit discussed in the case of the 4-point function \cite{Brower:2008nm}. Let us briefly review the preliminaries to consider such limit for our 5-point function: The first step is to define the set of momentum invariant quantities that we will use\footnote{For more details, see chapter 3 of \cite{Brower:2008nm}.}, consisting of two body energy invariants 
\begin{equation}
    s_{2} = -(p_{2}+p_{3})^{2}\,, \qquad s_{3}  = -(p_{3}+p_{4})^{2}\,,
\end{equation}
and two momentum transfers
\begin{equation}
    s_{1} = -(p_{1}+p_{2})^{2}\,, \qquad s_{4} = -(p_{4}+p_{5})^{2}\,.
\end{equation}
Momentarily, we are going to choose the three body energy
\begin{equation}
    s_{5} = -(p_{5}+p_{1})^{2}\,;
\end{equation}
and the convenient definition
\begin{equation}
    \kappa \equiv \dfrac{s_{5}}{s_{2}s_{3}}\,.
\end{equation}
Thus, the 5 independent momentum variables we are going to consider are\footnote{In general, the number of independent Lorentz invariant parameters in an amplitude should be $3n -10$.} $s_{1}$, $s_{2}$, $s_{3}$, $s_{4}$, and $\kappa$. The single-Regge limit corresponds to $s_{2}\rightarrow\infty$, 
$s_{5}\rightarrow\infty$, holding $s_{1}$, $s_{3}$, $s_{4}$, and $\kappa$ fixed, or equivalently
\begin{align}
    \dfrac{s_{1}}{s_{2}} & \rightarrow 0\,, & \dfrac{s_{3}}{s_{2}} & \rightarrow 0\,, & \dfrac{s_{4}}{s_{2}} & \rightarrow 0\,, & \kappa s_{2} \rightarrow \infty\,, \\
    \dfrac{s_{1}}{s_{5}} & \rightarrow 0\,, & \dfrac{s_{3}}{s_{5}} & \rightarrow 0\,, & \dfrac{s_{4}}{s_{5}} & \rightarrow 0\,, & \kappa s_{5} \rightarrow \infty\,.
\end{align}
This can conveniently be translated into the following set of collinear limits
\begin{equation}
    z_{12} \rightarrow \varepsilon\,, \qquad z_{34} \rightarrow \varepsilon\,, \qquad z_{45} \rightarrow \varepsilon\,,
    \label{eq:collinear-limits}
\end{equation}
with $\varepsilon \rightarrow 0$. We implicitly take the same limits for $\bz_{ij}$.
In such limits, the $f_{i5}$ functions become
\begin{align}
f_{15} & = \dfrac{-2\varepsilon}{z_{25}+\bz_{25}} + \mathcal{O}(\epsilon^{2})\,, \\
f_{25} & = \dfrac{2\varepsilon}{z_{25}+\bz_{25}} + \mathcal{O}(\varepsilon^{2})\,, \\
f_{35} & = 1 + 2 \left( \dfrac{1}{\bar{z}_{25}} + \dfrac{1}{z_{25}+\bz_{25}} \right)\varepsilon + \mathcal{O}(\varepsilon^{2})\,, \\
f_{45} & = -2 - \dfrac{2\varepsilon}{z_{25}+\bz_{25}} + \mathcal{O}(\varepsilon^{2})\,.
\end{align}

Hence, by taking the collinear limits in the integrand of \eqref{eq:5pointstring-st} the dominant contributions come from
\begin{equation}
    A  = -\dfrac{2\varepsilon}{z_{25}+\bz_{25}}|z_{25}|^{2}\log\left(\dfrac{x}{y}\right)\,, \qquad  c_1  = - \left( \dfrac{2\varepsilon}{z_{25}+\bz_{25}}|z_{25}|^{2}\right)^2\,,
\end{equation}
while $c_{2} \sim \mathcal{O}(\varepsilon^3)$ and  $c_{3} \sim \mathcal{O}(\varepsilon^4)$\footnote{The non-$\varepsilon$-dependent product, $\epsilon_{\alpha\beta\mu\nu}q_{2}^{\alpha}q_{2}^{\beta}q_{5}^{\mu}q_{5}^{\nu}$ becomes zero due to the antisymmetric behavior of the Levi-Civita symbol.}  do not contribute to the leading term.  Then, the expression of interest in equation \eqref{eq:5pointstring-st} is 
\begin{multline}
    -\dfrac{\alpha'^{\beta}}{2}\Gamma(2-\beta)\int_{0}^{1}dx \int_{0}^{1}dy\left(\dfrac{2\varepsilon}{z_{25}+\bz_{25}}|z_{25}|^{2}\right)^{2}\dfrac{1}{xy}\left[ -\dfrac{2\varepsilon}{z_{25}+\bz_{25}}|z_{25}|^{2}\log\left(\dfrac{x}{y}\right) \right]^{\beta-2}  \\
    = -\dfrac{\alpha'^{\beta}}{2}\Gamma(2-\beta)\left(\dfrac{2\varepsilon}{z_{25}+\bz_{25}}|z_{25}|^{2}\right)^{\beta}\int_{0}^{1}dx \int_{0}^{1}dy\dfrac{1}{xy}\left[ -\log\left(\dfrac{x}{y}\right) \right]^{\beta-2}\,.
    \label{string-integral-zij-limit}
\end{multline}

Focusing on the integral, we can easily see that it yields a $\delta$-function; in fact, by defining $u=-\log(x/y)$, we get
\begin{equation}
    \dfrac{1}{\beta-1}\int_{0}^{1}\dfrac{dx}{x}u^{\beta-1}\bigg\rvert_{-\infty}^{-\log x}\,,
\end{equation}
where the $u\rightarrow -\infty$ contribution vanishes because $\beta$ is a purely imaginary number, giving
\begin{equation}
    \dfrac{1}{\beta-1}\int_{0}^{1}\dfrac{dx}{x}\,(-\log(x))^{\beta-1}\,,
    \label{eq:integrand-zij-zero-limits}
\end{equation}
which is quite similar to the integral obtained in \cite{Stieberger:2018edy} for $n=4$, up to a factor $\beta-1$ in the denominator. As we will see, such factor only provides an overall minus sign. Defining $v=-\log(x)$, we have 
\begin{equation}
    \dfrac{1}{\beta-1}\int_{0}^{\infty}dv\, v^{\beta-1}\,,
\end{equation}
and, then, defining $\text{z}=\log(v)$,
\begin{equation}
    \dfrac{1}{\beta-1}\int_{-\infty}^{\infty} d\text{z}\, e^{\text{z}}\,e^{\text{z} \left(\beta -1\right)}\,.
\end{equation}
Last, recalling $\beta = -\frac{i}{2}\sum_{i=1}^{n}\lambda_{i} \equiv -\frac{i}{2}\lambda$, we may define $\tau =-z/2$ and recognize the well-known integral representation of the $\delta$-function
\begin{equation}
    \dfrac{2}{\beta-1}\int_{-\infty}^{\infty} d\tau \, e^{i\tau \lambda} = -4\pi\delta(\lambda)\,.
\end{equation}
By similarly applying the $\delta$-function in (\ref{string-integral-zij-limit}) and restoring all the factors of the amplitude in \eqref{eq:5pointstring-st}, we determine that the leading term of the 5-gluon superstring amplitude in the celestial basis yields
\begin{equation}
    \tilde{\mathcal{A}}_{I}^{\rm tree}(\lambda_{i},z_{i},\bz_{i}) = \dfrac{i}{4}\dfrac{(1-r_{4})(1-\bar{r}_{4})}{r_{4}-\bar{r}_{4}}\dfrac{z_{12}^{3}}{|z_{14}|^{2}|z_{23}|^{2}z_{23}z_{34}z_{45}z_{51}}f_{15}^{i\lambda_{1}+1}f_{25}^{i\lambda_{2}+1}f_{35}^{i\lambda_{3}-1}f_{45}^{i\lambda_{4}-1}2\pi\delta(\lambda)\,,\label{trenooo}
\end{equation}
which exactly reproduces the MHV 5-gluon YM amplitude in celestial basis. To see this in a more clear way, we can perform the Mellin transform of (\ref{5-point-aym}), in which only one of the $\omega_{i}$'s is not constrained by the $\delta $-functions. This yields
\begin{equation}
    \tilde{\mathcal{A}}_{\rm YM}^{\text{tree}}({-,-,+,+,+}) = \frac{i(1-r_4)(1-\bar{r}_4)}{4(r_4-\bar{r}_4)} \frac{f_{15}^{i\lambda_{1}+1}f_{25}^{i\lambda_{2}+1}f_{35}^{i\lambda_{3}-1}f_{45}^{i\lambda_{4}-1}\, z_{12}^3}{z_{23}z_{34}z_{45}z_{51}|z_{14}|^2|z_{23}|^2}\int_{0}^{\infty}d\omega_{5}\,\omega_{5}^{i\lambda-1}\,,\label{train}
\end{equation}
which, by defining $u=\log\omega_{5}$, gives the same integral representation of the $\delta $-function, so yielding (\ref{trenooo}).

\subsubsection{The \texorpdfstring{$z_{12}\to 0$}{} limit}

Now, let us examine a more general limit: consider the collinear limit $z_{12}\to 0$ of the celestial string 5-gluon amplitude, but keeping the other separations $z_{ij}$ finite. This can be regarded as a generalization of the single Regge limit considered above, and so it gives a more general picture of where in the ($z,\bar z$) plane the celestial YM observables are recovered. It is possible to show that the leading term in (\ref{eq:5pointstring-st}) is
\begin{equation}
    \dfrac{\alpha'^{\beta}}{2}\Gamma(2-\beta)\int_{0}^{1}dx\int_{0}^{1}dy\dfrac{{c}^{12}_{1}{c}^{12}_{2}}{xy}\left[ -{c}^{12}_{1}\log x - {c}^{12}_{2}\log y \right]^{\beta - 2}\,,
\end{equation}
where ${c}^{12}_{1}$ and ${c}^{12}_{2}$ depend on $z_{i\neq 1}$ and exhibit a $1/\epsilon$ factor arising when taking the collinear limit. We then proceed to integrate in $y$ by first defining $u=-(c_{1}^{12}/c_{2}^{12})\log x - \log y$. Integrating in $u$ and using properties of the $\Gamma$-function, we obtain
\begin{equation}
    -\dfrac{\alpha'^{\beta}}{2}\Gamma(1-\beta)(c^{12}_{1})^{\beta}\int_{0}^{1}\dfrac{dx}{x}\left( -\log x \right)^{\beta-1}\,,
\end{equation}
which is the same integral as in (\ref{eq:integrand-zij-zero-limits}), leading to the same expression, including the $\delta$-function. After evaluating the $\delta$-function, and reinstating the additional factors, we recover the MHV 5-gluon YM amplitude in celestial basis. In other words, we have managed to identify the limit in which the field theory results are obtained in the celestial basis.

\subsubsection{Collinear limits}

It is important not to mistake the specific limits considered above for the generic collinear limits. The latter do not necessarily lead to the field theory result. This is analog to what happens in the case $n=4$, where the forward scattering limit is actually the one reproducing the YM celestial amplitudes. In the case, $n=5$ something analogous occurs: in contrast to the limits $z_{12}\to 0$ discussed above, other coincident limits $z_{ij}\to 0$ in the 5-gluon celestial string amplitude do not lead to the same integral expression obtained in (\ref{eq:integrand-zij-zero-limits}). In order to make it clear, we find convenient to write the following table displaying the general form that the $x$- and $y$-dependent integrand takes in each limit:
\newcommand\T{\rule{0pt}{4ex}}       
\newcommand\B{\rule[-2.6ex]{0pt}{0pt}} 
\begin{table}[H]
\centering
\begin{longtable}{|c|c|} 
\hline 
$z_{12}\rightarrow 0$ & $\dfrac{c_{1}}{xy}\left(c^{12}_{1}\log x + c^{12}_{2}\log y \right)^{\beta-2}$ \T\B \\ 
$z_{13}\rightarrow 0$ & $\left(\dfrac{c_{1}}{xy} + \dfrac{c_{2}}{1-xy}\right)\left(c^{13}_{1}\log x + c^{13}_{2}\log y + c^{13}_{3}\log(1-x) + c^{13}_{5}\log(1-xy) \right)^{\beta-2}$ \T\B \\ 
$z_{14}\rightarrow 0$ & $\left(\dfrac{c_{1}}{xy} + \dfrac{c_{2}-2ic_{3}}{1-xy}\right)\left(c^{14}_{2}\log y + c^{14}_{3}\log(1-x) + c^{14}_{4}\log(1-y) \right)^{\beta-2}$ \T\B \\ 
$z_{15}\rightarrow 0$ & $\dfrac{c_{2}-2ic_{3}}{1-xy}\left(c^{15}_{4}\log(1-y) + c^{15}_{5}\log(1-xy)\right)^{\beta-2}$ \T\B \\ 
$z_{23}\rightarrow 0$ & $\dfrac{c_{2}-2ic_{3}}{1-xy}\left(c^{23}_{3}\log(1-x) + c^{23}_{5}\log(1-xy)\right)^{\beta-2}$ \T\B \\
$z_{24}\rightarrow 0$ & $\left(\dfrac{c_{1}}{xy} + \dfrac{c_{2}-2ic_{3}}{1-xy}\right)\left(c^{24}_{1}\log x + c^{24}_{3}\log(1-x) + c^{24}_{4}\log(1-y)\right)^{\beta-2}$ \T\B \\
$z_{25}\rightarrow 0$ & $\left(\dfrac{c_{1}}{xy} + \dfrac{c_{2}}{1-xy}\right)\left(c^{25}_{1}\log x + c^{25}_{2}\log y + c^{25}_{4}\log(1-y) + c^{25}_{5}\log(1-xy)\right)^{\beta-2}$ \T\B \\
$z_{34}\rightarrow 0$ & $\left(\dfrac{c_{1}}{xy} + \dfrac{c_{2}-2ic_{3}}{1-xy}\right)\left(c^{34}_{1}\log x + c^{34}_{4}\log(1-y) + c^{34}_{5}\log(1-xy)\right)^{\beta-2}$ \T\B \\ 
$z_{35}\rightarrow 0$ & $\dfrac{c_{2}-2ic_{3}}{1-xy}\left(c^{35}_{1}\log x + c^{35}_{2}\log y + c^{35}_{3}\log(1-x) + c^{35}_{4}\log(1-y)\right)^{\beta-2}$ \T\B \\ 
$z_{45}\rightarrow 0$ & $\left(\dfrac{c_{1}}{xy} + \dfrac{c_{2}-2ic_{3}}{1-xy}\right)\left(c^{45}_{2}\log y + c^{45}_{3}\log(1-x) + c^{45}_{5}\log(1-xy)\right)^{\beta-2}$ \T\B \\ \hline
\end{longtable}
\end{table}

The coefficients $c^{ij}_k$, $i,j,k=1,...,5$, can be written explicitly and turn out to be independent of $z_{j}$. It would be interesting to further investigate the hierarchical collinear limits of the string $n$-gluon amplitudes in the celestial basis and use them to explore the different limits in which YM celestial amplitudes are reproduced. To that purpose, it would also be interesting to extend the analysis to the specific cases $n=6$ and $n=7$.

\section{Conformally soft theorem}

In this section we would like to discuss the conformally soft theorems \cite{Adamo:2019ipt,Nandan:2019jas,Fan:2019emx,Pate:2019mfs, Donnay:2018neh,Puhm:2019zbl} for the case of the 5-gluon string amplitude.

Let us start by first reviewing the energetic soft theorem for the case of 5-gluons in the type-I theory at tree level, as originally found by Stieberger and Taylor \cite{Stieberger:2006bh,Stieberger:2006te}. The 5-gluon amplitude is given by \eqref{5gluon-string}, with $P^{(5)}$ and $V^{(5)}$ given by \eqref{eq:p5} and \eqref{eq:v5}. Now, consider the case in which the fifth particle goes soft, {\it i.e.} take $p_5 \to 0$. In this limit, one has
\begin{align}
P^{(5)}(s_i) & =  \mathcal{O}(1) \, ,\\
V^{(5)}(s_i) & =  \frac{\Gamma(1+\alpha' s)\Gamma(1+\alpha't)}{\Gamma(1+\alpha's +\alpha't)} + \mathcal{O}(p_5)\, ,
\\
\epsilon(1,2,3,4) & = \mathcal{O}(p_5)\, ;
\end{align}
therefore,
\eal{
\label{eq:soft-5-point-string}
\lim_{p_5 \to 0} \mathcal{A}_I^{\rm string} (-,-,+,+,+) = \frac{\Gamma(1+\alpha' s)\Gamma(1+\alpha't)}{\Gamma(1+\alpha's +\alpha't)} \lim_{p_5 \to 0} \mathcal{A}_{I}^{\rm YM}(-,-,+,+,+)\,,}
leading to
\eal{
\label{eq:soft-5-point-string2}
\lim_{p_5 \to 0} \mathcal{A}_I^{\rm string} (-,-,+,+,+) \, =\, S^{(0)}\mathcal{A}_I^{\rm string}(-,-,+,+)\,,
}
where $S^{(0)}$ is the leading, tree-level soft factor for YM theory; notice that, in (\ref{eq:soft-5-point-string2}), the left-hand side involves a 5-point string amplitude while on the right-hand side involves a 4-point string amplitude. Thus, we see that the string amplitude also satisfies the soft theorem with precisely the same universal soft factor that appears for Yang-Mills theory, $S^{(0)}$. With this in hand, it is now possible to show that the conformally soft theorem for the string celestial amplitude is also obeyed. 

Let us first review the general result for the conformally soft limit in gauge theory \cite{Adamo:2019ipt,Nandan:2019jas,Fan:2019emx,Pate:2019mfs}. Consider the case when the $k^{\text{th}}$ gluon in a (color-ordered) celestial amplitude goes conformally soft, corresponding to $\Delta_k \to 1$, {\it i.e.}
\eal{
\lim_{\lambda_k \to 0} \tilde{\mathcal{A}}_n(\{\Delta_i\})&= \lim_{\lambda_k \to 0} \left( \prod_{j\neq k}^n \int_0^{\infty} d\omega_j \, \omega_j^{i\lambda_j}\right)  \int_0^{\infty} d\omega_k \, \omega_k^{i\lambda_k} \mathcal{A}_n(\{p_i\})\,.
}
Here, $\tilde{\mathcal{A}}_n(\{\Delta_i\})$ and $\mathcal{A}_n(\{p_i\})$ stand for the $n$-point amplitudes in the conformal and momentum basis, respectively. Using the representation of the Dirac $\delta$-function
\eal{
\delta(x) = \lim_{\lambda \to 0} \frac{i\lambda}{2} |x|^{i\lambda-1} \,,
}
one can write
\eal{
\label{eq:conformally-soft-1}
\lim_{\lambda_k \to 0} \tilde{\mathcal{A}}_n(\{\Delta_i\})&= \lim_{\lambda_k \to 0} \, \frac{2}{i\lambda_k} \left( \prod_{j\neq k}^n \int_0^{\infty} d\omega_j \, \omega_j^{i\lambda_j}\right)  \int_0^{\infty} d\omega_k \, \frac{i\lambda_k}{2}\omega_k^{i\lambda_k-1} \omega_k \mathcal{A}_n(\{p_i\})\\
&=\lim_{\lambda_k \to 0} \, \frac{2}{i\lambda_k} \left( \prod_{j\neq k}^n \int_0^{\infty} d\omega_j \, \omega_j^{i\lambda_j}\right)  \int_0^{\infty} d\omega_k \, \delta(\omega_k) \omega_k \mathcal{A}_n(\{p_i\})\,.
}
In the last line we see that the effect of the $\delta$-function is to select the soft limit $\omega_k \to 0$ of the momentum-space amplitude ${A}_n(\{p_i\})$, which we already know from its energetic soft theorems. More precisely, at tree level, the single soft gluon theorem for the amplitude reads
\eal{
\label{eq:soft-gluon}
\mathcal{A}_n = S^{(0)}\mathcal{A}_{n-1}+\mathcal{O}(\omega_k^0)\,,
}
with
\eal{
\label{eq:soft-factor}
S^{(0)} = \frac{1}{\sqrt{2}}\left(\frac{\epsilon_k\cdot p_{k+1}}{p_{k+1}\cdot p_k}-\frac{\epsilon_k\cdot p_{k-1}}{p_{k-1}\cdot p_k}\right) = \frac{1}{\sqrt{2}\,\eta_k}\left(\frac{\epsilon_k\cdot q_{k+1}}{q_k\cdot q_{k+1}}-\frac{\epsilon_k\cdot q_{k-1}}{q_k\cdot q_{k-1}}\right)\frac{1}{\omega_k}\,,
}
where, in the second equality above, we have again used the usual parametrization $p_i^{\mu}=\eta_i \omega_i q_i^{\mu}$ and $\epsilon_k$ is the polarization vector of the $k^{\text{th}}$ gluon; also, we have denoted $\epsilon_k\cdot q_{k'}=\epsilon_{k\, \mu }\, q^{\mu }_{k'}$. Notice that all the $\omega_k$ dependence in the soft limit \eqref{eq:soft-gluon} lies in the soft factor \eqref{eq:soft-factor} since $\mathcal{A}_{n-1}$ is the hard amplitude with the $k^{\text{th}}$ particle removed. Using this and inserting \eqref{eq:soft-gluon} and \eqref{eq:soft-factor} into \eqref{eq:conformally-soft-1} allows us to explicitly solve for the integral over $\omega_k$ yielding
\begin{multline}
\label{eq:conformally-soft-2}
\lim_{\lambda_k \to 0} \tilde{\mathcal{A}}_n(\{\Delta_i\})
=\lim_{\lambda_k \to 0} \, \frac{2}{i\lambda_k} \left( \prod_{j\neq k}^n \int_0^{\infty} d\omega_j \, \omega_j^{i\lambda_j}\right)  \\
\times \frac{1}{\sqrt{2}\,\eta_k}\left(\frac{\epsilon_k\cdot q_{k+1}}{q_k\cdot q_{k+1}}-\frac{\epsilon_k\cdot q_{k-1}}{q_k\cdot q_{k-1}}\right)\mathcal{A}_{n-1}\int_0^{\infty } d\omega_k \, \delta(\omega_k) \,,
\end{multline}
obtaining
\eal{
\lim_{\lambda_k \to 0} \tilde{\mathcal{A}}_n(\{\Delta_i\})
=  \frac{1}{\sqrt{2}\,\eta_k}\left(\frac{\epsilon_k\cdot q_{k+1}}{q_k\cdot q_{k+1}}-\frac{\epsilon_k\cdot q_{k-1}}{q_k\cdot q_{k-1}}\right)\frac{1}{i\lambda_k} \tilde{\mathcal{A}}_{n-1} + \mathcal{O}(\lambda_k^0)\,,
}
{\it i.e.}, in the {\it conformally soft} limit $\lambda_k \to 0$ we again have the factorization into a universal (conformal) soft factor $\tilde{S}^{(0)}$ and the {\it conformally hard} $(n\!-\!1)$-point celestial amplitude $\tilde{\mathcal{A}}_{n-1}$ with
\eal{
\tilde{S}^{(0)} = \frac{1}{\sqrt{2}\,\eta_k}\left(\frac{\epsilon_k\cdot q_{k+1}}{q_k\cdot q_{k+1}}-\frac{\epsilon_k\cdot q_{k-1}}{q_k\cdot q_{k-1}}\right)\frac{1}{i\lambda_k}\,.
}
With this and \eqref{eq:soft-5-point-string}, we can readily apply this to the string amplitudes. That is, we write down
the conformally soft theorem for the celestial 5-gluon amplitude in the type I superstring theory as follows
\eal{
\tilde{\mathcal{A}}_I^{\rm string}(-,-,+,+,+)
=  \frac{1}{i\sqrt{2}\, \eta_5\lambda_5}\left(\frac{\epsilon_5\cdot q_{1}}{q_5\cdot q_{1}}-\frac{\epsilon_5\cdot q_{4}}{q_5\cdot q_{4}}\right) \tilde{\mathcal{A}}_{I}^{\rm string} (-,-,+,+) +\mathcal{O}(\lambda_5^0)\,,
}
which relates $5$-point string amplitudes, in the conformally soft limit, to $4$-point string amplitudes in the celestial basis. Here, we have also made use of the fact that, for color-ordered $n$-gluon amplitudes, we have the cyclic identification $k\!+\!n \sim k$.

\section{Mapping the string worldsheet to the CCFT}

Before concluding, let us move to discuss another interesting feature of the celestial string amplitudes. In \cite{Stieberger:2018edy} Stieberger and Taylor made a very interesting observation regarding the high-energy limit of the tree level 4-gluon string amplitude and its corresponding celestial correlator, which we briefly review now.

At tree level, the $SL(2,\mathbb{R})$ invariance of the moduli integrals, representing an $n$-point open string amplitude, allows one to fix the location of three out of the $n$ vertex operators to any desired value.\footnote{Therefore, the full tree-level $n$-point open string amplitude involves integrating over $n\!-\!3$ real moduli.} Thus, in the 4-gluon case, only one (real) integration variable is needed to write the full amplitude (see, \emph{e.g.}, \eqref{eq:4-point-integral}). Using this integral representation we can use the stationary-point approximation to evaluate the integral in the high-energy limit at fixed scattering angle.\footnote{This regime is also known as the \emph{hard scattering} limit.} This procedure yields one constraint that fixes the location of the vertex operator which, in turn, becomes completely determined by the angular position of the external momenta on the celestial sphere. More precisely, the stationary-point equation ties the worldsheet location of this vertex operator to the kinematic ratio $r$ \eqref{eq:cross-ratio}, thus identifying a point on the worldsheet with a point on the celestial sphere (see section 4 in \cite{Stieberger:2018edy} for more details). 

Recall that, in celestial amplitudes, the Mellin transform involves integrating over all of the energies in the scattering process, thus, erasing the usual hierarchy among the low energy and the high energy regimes. After computing these energy integrals, celestial string amplitudes thus remain written as integrals over the moduli, as in \eqref{Iintegral} for instance. This raises the question as to whether there is also a limit, now based on celestial data only, in which this \emph{pinning} on the celestial sphere also occurs in the celestial basis. In \cite{Stieberger:2018edy} Stieberger and Taylor show that this is indeed the case, and that the limit in which it occurs is the limit of large (imaginary) total scaling dimension, \emph{i.e.}, when  $\sum_{i=1}^4 \lambda_i \equiv \lambda \to \infty$.

Due to this result, we would like to pursue the idea of a potential map between the string worldsheet, punctured by the insertions of vertex operators, and the celestial sphere, also punctured but by the conformal primary operators corresponding to each of the external states with momentum $p_i$ that these vertex operators represent. Thus, the natural next step would be to extend this analysis to five points and beyond, which is what we study in this section.

For the 5-gluon case in string theory, let us first re-write the integrands for the amplitude in \eqref{5gluon-string} as
\begin{equation}
    P^{(5)}(s_i) = \int^{1}_{0}\int^{1}_{0}\,dx\, dy\, (1-xy)^{-1}\, e^{\alpha' f(x,y)}\,,
\end{equation}
and
\begin{multline}
      V^{(5)}(s_i) = \tfrac{1}{2}\alpha'^2\left(s_2s_3+s_4s_5-s_1s_2-s_3s_4-s_1 s_5\right)P^{(5)}(s_i)\\
      +\alpha'^2s_2s_5 \,\int^{1}_{0}\int^{1}_{0}\,dx dy\, (xy)^{-1}\, e^{\alpha' f(x,y)}\,, \hspace{50pt}
\end{multline}
with
\eal{
f(x,y)&=s_2 \ln x +s_5\ln y + s_3 \ln(1-x)+s_4\ln(1-y) + (s_1-s_3-s_4)\ln(1-xy)\,.
}
Before continuing, it is worth mentioning a general feature of string amplitudes. Both integrals above will give convergent expressions as long as the Mandelstam invariants ${s_i}$ are constrained to certain regions in the complex plane.
These regions will not necessarily lie in the domain of physical scattering, but they allow us to compute the amplitude and also to use the stationary-point approximation to evaluate it in the high energy limit. Nevertheless, one usually computes these integrals in the (unphysical) convergent regions and, at the end, we analytically continue the resultant expressions to the physical domain.\footnote{See, \emph{e.g.}, section 6.4 in \cite{Polchinski:1998rq} for a discussion on these points for tree level amplitudes, and \cite{Thorn:2008ay,Rojas:2011wx,Rojas:2013oca} for the extension to one-loop.}

In the hard scattering limit, namely when $\alpha' s_i \gg 1$ with all the Mandelstam ratios $s_i/s_j$ held fixed, we can now evaluate the integrals above using the stationary-point method. The stationary-point equations 
$\partial_x f = \partial_y f = 0$ now yield two (independent) solutions, say, $(x_1,y_1)$ and $(x_2,y_2)$. The exact expressions for these two solutions can be straightforwardly written but are cumbersome and not very illuminating. However, the relevant observation is that, since the $x$ and $y$ variables represent the positions of the two remaining vertex operators that need to be integrated over to obtain the full 5-point amplitude, the high-energy limit yields now two possible configurations that give the dominant contributions in this limit.\footnote{In contrast with the 4-point case in which there is only one dominant saddle.} The explicit expressions for these two solutions, when written using the usual celestial parametrization $p_i^{\mu}=\eta_i \omega_i q_i^{\mu}(z_i,\bar{z}_i)$, show that they depend solely on the angular positions of the external states $(z_i,\bar{z}_i$) on the celestial sphere. This is because total momentum conservation makes all the energy factors $\omega_i$ to cancel out in these solutions. Thus, in the 5-gluon case, we can again interpret this as the fact that the vertex insertions on the string worldsheet become \emph{pinned} by the locations of their corresponding primaries of the celestial CFT. However, there is a caveat: we now have two possible configurations of vertex operators on the worldsheet corresponding to a single configuration on the celestial sphere, that is, the map is no longer bijective.

Now, similarly to the 4-point case, the 5-gluon celestial string amplitude \eqref{eq:5pointstring-st} is also manifestly written in terms of the moduli space integrals, inhereted from its momentum space counterpart. Thus, we can ask ourselves again in which limit does the celestial amplitude localize the vertex operators to the same positions that dominate the high-energy regime. From \eqref{mellin-string-A} and \eqref{eq:5pointstring-st} one can see that, again, it is the $\sum_{i=1}^5 \lambda_i \equiv \lambda \to \infty$ limit that yields the same stationary-point equations as in momentum space.

These results, therefore, prompt us to understand how to extend this analysis to the arbitrary $n$-point string amplitude. The sought connection between the string worldsheet and the celestial sphere seems, however, to be clearer in the case of closed strings for which, at tree level, the worldsheet topology is also a sphere. 

Let us start by writing down the expression representing a tree-level, $n$-point, closed string amplitude. This amplitude takes the general form\footnote{We use here $w_k$ to denote the complex coordinates on the string worldsheet in order to distinguish them from the $z_i$ parametrizing the angular directions of the external states on the celestial sphere.}
\eal{
\label{eq:n-closed}
\int \left(\prod_w d^2 w_k\right) \mathcal{P}(\{w_i\},\{p_i \cdot p_j\}) \prod_{i<j}  |w_j-w_i|^{2\alpha' p_j \cdot p_i}\,,
}
where $\mathcal{P}$ is a polynomial function of the Mandelstam products $\alpha' p_i \cdot p_j$ whose degree will depend on the spin of the scattering external states. More importantly, the $|w_j-w_i|^{2\alpha' p_j \cdot p_i}$ factor is universal to all string amplitudes\footnote{This is because this factor comes from the Wick contractions of vertex operators $V(p_i)$ which all contain the plane-wave factor $V(p_i) \sim e^{i\,  p_i \cdot X}$ representing free asymptotic states with definite momentum.} and depends exponentially on the Mandelstam products $\alpha' p_j \cdot p_i$, thus being responsible for yielding the dominant contributions in the high-energy limit $\alpha' p_i \cdot p_j \gg 1$. With this, we can use again the saddle-point method to obtain the leading contribution in this limit. From \eqref{eq:n-closed}, one can readily see the saddle-point equations are \cite{Gross:1987ar,Gross:1987kza}
\eal{
\label{eq:scattering-equations}
\sum_{j\neq i} \frac{p_i \cdot p_j}{w_i-w_j}=0 \,, \qquad i=1,\dots, n.
}
These equations have received significant attention relatively recently due to their key role in the Cachazo-He-Yuan (CHY) formulation of scattering amplitudes for massless particles \cite{Cachazo:2013iea,Cachazo:2013hca}, and are usually referred to as the \emph{scattering equations}. After imposing total momentum conservation, and the massless condition $p_i^2=0$, the equations \eqref{eq:scattering-equations} are $SL(2,\mathbb{C})$ invariant and thus, permit to fix three of the $n$ positions $\{w_i\}$ to any convenient value on the worldsheet. The number of independent solutions, \emph{i.e.}, the number of distinct possible configurations for the positions of the vertex operators on the worldsheet, is $(n-3)!$ \cite{Cachazo:2013gna}. Moreover, using elimination theory, the system of equations \eqref{eq:scattering-equations} can be brought into the form of a single polynomial equation of degree $(n-3)!$ for a single variable, say, $w_n$ \cite{Cardona:2015ouc,Dolan:2015iln}. The solutions for all the remaining vertex locations $w_i$, with $i=4,\dots, (n-1)$, can be expressed in terms of the single solution $z_n$ and become functions of ratios of generalized Mandelstam variables. After using total momentum conservation, the positions of all these punctures will again only depend on the angular positions $z_i$ of the scattered strings on the celestial sphere. 

This establishes, then, the general map that emerges in the high-energy limit: A single configuration on the celestial sphere, for the momenta of $n$ closed strings in a (tree level) amplitude, gets mapped to $(n-3)!$ possible configurations of vertex operators on the string worldsheet that govern the scattering process in this regime. It would certainly be interesting to confirm if one arrives at the same conclusion by taking $\sum_{i=1}^n \lambda_i \to \infty$ at the level of the $n$-point celestial amplitude. 

As a final point, it is also interesting to note that very same configuration on the celestial sphere (at null infinity), representing the insertions of conformal primaries for each of the scattered gravitons,\footnote{The same statement certainly holds for the other states in the massless spectrum of the closed string, such as the dilaton and the Kalb-Ramond field.} also solves the saddle-point equations for the corresponding vertex operators on the worldsheet sphere \eqref{eq:scattering-equations}. Namely, from the usual parametrization for the external momenta 
$$p^\mu_i = \tfrac{1}{2}\eta_i \omega_i (1+|z_i|^2, z_i+\bar{z}_i,-(z_i-\bar{z}_i),1-|z_i|^2),$$
one has $p_i\cdot p_j = \eta_i \eta_j \omega_i \omega_j |z_{ij}|^2 $, thus, if $z_k=w_k$ for all $k=1,\dots, n$, we have
\eal{
\sum_{j\neq i}^n \frac{\eta_i \eta_j \omega_i \omega_jw_{ij}\bar{w}_{ij}}{w_{ij}}&=\eta_i \omega_i\left(\bar{w_i}\sum_{j\neq i}^n \eta_j \omega_j -\sum_{j\neq i}^n \eta_j \omega_j \bar{w_j}\right)\\
&=\eta_i \omega_i\left(-\bar{w_i}\eta_i \omega_i +\eta_i \omega_i \bar{w_i}\right)\\&=0,
}
where, in going from the first to the second line, we have used total momentum conservation. Thus, the configuration of physical (stringy) gravitational scattering on the celestial CFT is one of the configurations of this worldsheet map that arises in the high-energy limit. It would be very interesting to pursue this idea further.
  
\section{Concluding remarks}

In this last section we would like to make a brief observation regarding the effective field theory expansion of string amplitudes in the celestial basis, that is, we study how the $\alpha'$ expansion is recast in the conformal basis. After this, we finalize with some closing words.

Consider the tree-level string amplitude of 4 gluons written in the form that makes the connection with YM amplitudes evident in the $\alpha' \to 0$ limit, namely
\eal{
{\mathcal {A}}_I^{\text{tree}}(-,-,+,+)=\delta^{(4)}\left(\sum_{i=1}^4 p_i^{\mu}\right) \frac{\Gamma(1-\alpha' s)\Gamma(1-\alpha't)}{\Gamma(1-\alpha's -\alpha't)}{\mathcal {A}}_{\text{YM}}^{\text{tree}}(-,-,+,+)\,,
}
and let us expand it in powers of $\alpha'$
\begin{multline}
{\mathcal {A}}_I^{\text{tree}}(-,-,+,+)=\delta^{(4)}\left(\sum_{i=1}^4 p_i^{\mu}\right)
{\mathcal {A}}_{\text{YM}}^{\text{tree}}(-,-,+,+) \left(1-\alpha'^2 \frac{\pi^2}{6}st-\alpha'^3 \zeta(3)(s^2 t+s t^2)\right.\\-\left.\alpha'^4\frac{\pi^4}{360}(4 s^3 t+s^2 t^2 +4 s t^3)+\cdots\right)\,.
\end{multline}
The celestial counterpart of this expression reads
\eal{
\tilde{\mathcal {A}}_I^{\text{tree}}(-,-,+,+)= f(r,\beta) \prod_{i<j}^4 z_{ij}^{h/3-h_i-h_j}\bar{z}_{ij}^{\bar{h}/3-\bar{h}_i-\bar{h}_j}\,,
}
with $f(r,\beta)$ now given by
\begin{multline}
f(r,\beta) = \int_0^{\infty} dw \, w^{-\beta-1} \left[1+\alpha'^2 \frac{\pi^2}{6}rw^2+\alpha'^3 \zeta(3)(r^2-r)w^3\right.\\+\left.\alpha'^4\frac{\pi^4}{360}(4 r^3 - r^2 +4 r)w^4+\cdots\right]\,,
\end{multline}
up to some, for now, irrelevant numerical factors. Notice that, since $\beta$ is purely imaginary, the integrals over all terms are UV divergent, except for the $1$-term which (marginally) converges to the known $\delta$-function $4\pi \delta(\sum_{i}\lambda_i)$ appearing in the field theory limit. Thus, in order to make sense of the $\alpha'$ expansion in the celestial basis, we need to introduce an UV cutoff on the energy integral above.\footnote{This is another example of the fact that celestial amplitudes are usually not well-defined for effective field theories and need an UV completion. However, see \cite{Mitra:2024ugt} for a very recent proposal on how to define them in a consistent way. Additionally, see \cite{Ren:2022sws} for a detailed study on the constraints that effective field theories with celestial duals must satisfy, even at tree level.} Let us call the cutoff $\Lambda$, with $\Lambda$ being the energy scale. The full expression for $f(r,\beta)$ can then be written as
\begin{multline}
\label{f-expansion}
f(r,\beta) = 4\pi \delta(\lambda)+ \frac{r}{\Lambda^\beta} (\alpha'\Lambda^2)^2\left[\frac{\pi^2}{6}\frac{1}{(2-\beta)} - 
(\alpha' \Lambda^2)\zeta(3)(1-r)\frac{1}{(3-\beta)}\right.\\
-
\left.(\alpha' \Lambda^2)^2\frac{\pi^4}{360}(-4r^2+r-4)\frac{1}{(4-\beta)}+\cdots\right]\,,
\end{multline}
where $\lambda = \sum_{i=1}^4 \lambda_i$. From this, we see that the dimensionless parameter that organizes the effective field theory expansion is $\alpha' \Lambda^2$, which was of course expected on dimensional grounds since these are the only two dimensionful quantities at our disposal. Notice also the appearance of the pole structure in the $\beta$-complex plane. As initially analyzed in \cite{Arkani-Hamed:2020gyp} in the context of effective field theories with exponentially soft UV behavior, the residues of the poles in \eqref{f-expansion} are precisely given by the coefficients obtained by the low-energy expansion of the original string theory amplitude.

Let us summarize the results of this work as follows: In this paper, we have computed celestial correlators that correspond to the tree-level 5-gluon amplitudes in type I superstring theory. We have observed that, as it happens with the 4-gluon string celestial amplitudes, both at tree-level \cite{Stieberger:2018edy} and at one-loop \cite{Donnay:2023kvm}, the 5-gluon string celestial amplitudes exhibit a remarkably simple dependence of $\alpha '$, which is encapsulated in a universal overall factor. This raised the question as to how the field theory limit is recuperated from the string observables in the celestial basis. This motivated our study of the limit in the space of kinematic variables where the YM results are recovered. We have identified such limit, generalizing in this way the results obtained by Stieberger and Taylor in \cite{Stieberger:2018edy}. We have also computed the conformally soft theorem for the 5-gluon case in the type I superstring theory and confirmed that it also obeyed as in the case of celestial field theory amplitudes. This was expected, at least at the tree-level we have considered here since, due to the fact that field theory arises in the low energy limit of strings, the soft factorization for both theories should be identical, both in momentum and conformal basis. Finally, we discussed the map that emerges between the puctured string worldsheet CFT and the also punctured celestial CFT, in the limit of high-energy, in momentum basis, corresponding to high-scaling-dimension, in conformal basis, for a general $n$-point string amplitude. In order to gain more insights into this correspondence, it would be interesting to perform the explicit computations to higher-point amplitudes. We hope to address this and other questions in the near future.

\section*{Acknowledgments}
The authors would like to thank Laura Donnay, Hernán González, Chrysostomos Kalousios, Lorenzo Magnea, Andrea Puhm, Ana-Maria Raclariu, and Tomasz Taylor for  helpful discussions. They also thank Anthony Osses for collaboration during the first stages of this work. The work of L.C. is supported by the STFC Astronomy Theory Consolidated Grant ST/W001020/1 from UK Research \& Innovation. G.M. has been supported by ANID/ACT 210100 Anillo grant, and F.R. has been supported by FONDECYT grant 1221920 and ANID/ACT 210100 Anillo grant.

\bibliography{celestial-refs}
\bibliographystyle{utphys}

\end{document}